\documentclass[12pt]{article}
\usepackage{amsmath,amsfonts,amssymb}
 \usepackage{epsf,graphics,graphicx}

\newcommand{\be}{\begin{equation}}

\newcommand{\ee}{\end{equation}}
\newcommand{\bea}{\begin{eqnarray}}
\newcommand{\eea}{\end{eqnarray}}
\newcommand{\bee}{\begin{equation*}}
\newcommand{\eee}{\end{equation*}}

\newcommand{\nn}{\nonumber}

\newcommand{\myfig}[4]{
\begin{figure}[h]
\begin{center}
\includegraphics[scale=#2]{#1}
\end{center}
\caption{#3}
\label{fig:#4}
\end{figure}
}

\usepackage{color}

\begin{document}

\title{Optimal Trading with Alpha Predictors}
\author{  Filippo Passerini\footnote{filippop@princeton.edu}  $\,\,$and    Samuel E. V\'azquez\footnote{svazquez@vegaedge.com}   \\
\small  ${}^\ast$Department of Physics, Princeton University, Princeton, NJ 08544\\
\small Laboratoire de Physique Th\'{e}orique, Ecole Normale Sup\'{e}rieure, 75005 Paris, France\\
\small ${}^\dagger$Vega Edge LLC, 48 Wall Street 11th Floor, New York, NY 10005
\\
\\
}

\maketitle

\abstract{
We study the problem of optimal trading using general alpha predictors with linear costs and temporary impact.
We do this within the framework of stochastic optimization with finite horizon using both limit and market orders. 
Consistently with other studies, we find that the presence of linear costs induces a ``no-trading" zone when using market orders, and a corresponding ``market-making" zone when using limit orders. We show that, when combining both market and limit orders, the problem is further divided into zones in which we trade more aggressively using market orders. Even though we do not solve analytically the full optimization problem, we present explicit and simple analytical ``recipes" which approximate the full solution and are easy to implement in practice. We test the algorithms using Monte Carlo simulations and show how they improve our Profit and Losses. 
}

\section{Introduction}
In this paper we take the point of view of a general trader who wants to profit from all possible sources of alpha, including predictors and even possibly market-making with limit orders. Hence, we present a very general and practical framework inspired by the Hamilton-Jacobi-Bellman (HJB) method.
We do not seek to find exact analytical solutions to the optimization problems, but to introduce some algorithmic ``recipes" which can be easily implemented in real trading. We allow the use of both limit and market orders and general alpha predictors. We show that many popular optimization problems in the literature can be viewed as special cases of our framework.

The study of optimal trading under linear costs has a long history. We will not review all the literature here but instead point the reader to the following review: \cite{review}. Previous works have mostly  focused on the case of optimal investment, where the agent has the choice of a risky asset and a savings account, or in the case of optimal consumption (see e.g. \cite{davis, shreve}). Moreover, the most popular set up is the stochastic control optimization with infinite time horizon. These works have all confirmed that the optimal strategy involves trading instantaneously to the boundary of a no-trading (NT) zone centered at the Markowitz Portfolio. As the name suggests, inside this zone, the agent does not trade. The width of the zone has been estimated by many authors using different approximation methods (see e.g. \cite{soner,kallsen,bichuch}). However, no closed-form solutions are known. In fact, the  full optimization problem is quite intractable. The majority of the literature on linear costs does not consider the predictability of the risky asset (apart from a drift). The notable exceptions are: \cite{cfm,martin,martin1}. In these papers the explicit width of the NT zone is estimated in the limit of small linear costs and mean-reverting predictors. 

Another source of slippage is price impact, see e.g. \cite{impact}. This effect is only important when we trade a non-trivial fraction of the daily volume. It is well known that impact can be divided into a permanent and temporary part. Empirically, it is found that  temporary impact leads to an aggregated cost which is a power of the total transacted quantity $Q$: $C(Q) \sim Q^{3/2}$. For mathematical simplicity, many papers on optimization simplify this power-law behavior with a simpler quadratic cost model: $C(Q) \sim Q^2$. In fact, the problem of trading mean-reverting alpha predictors with quadratic costs can be solved in closed form \cite{garlenau}. The optimal liquidation problem under temporary impact has recently been studied in \cite{curato}. The optimal investment problem has been studied under the presence of both linear and quadratic costs in \cite{liu}. 

Another rather separated stream of literature has studied the use of limit orders both for the liquidation \cite{lehalle1,chevalier1} and market-making \cite{avellaneda}-\cite{guilbaud1}. The problem of optimally placing market versus limit orders under the possibility of partial execution was studied in \cite{cont,guo, huitema}. For market-making, the main problem is the optimal price of the limit orders and the inventory risk. If we ignore partial execution, one can obtain explicit solutions for the HJB equations \cite{lehalle1,lehalle}. One piece missing from most of these articles is the addition of general alpha predictors, as the focus has been mainly market-making or liquidation. In \cite{guilbaud1} some basic predictability was introduced in the price process. Another problem with limit orders is that the optimization problems can be even more intractable that in the case of linear costs (market orders). 

In this paper we seek to bring together ingredients from all these works to formulate an optimization problem for a general alpha trader. We do this with a finite time horizon. This choice of boundary condition is motivated by the reality of trading. With the exception of FX markets, most assets are traded over finite market hours and marked-to-market to the close of the trading day. In fact, the standard accounting practice is to divide the total daily P\&L into the part coming from the trades (trade-to-close) and the part coming from the open positions (close-to-close). As traders, we can only control the trading during the present day and possibly the overnight risk we want to keep for the next day. This is specially important for high-frequency traders and market makers as they must limit the amount of overnight risk. Another important fact about trading is that often we use signals that are calculated with daily data which is officially available only after the market closing time. This might be for operational reasons or due to the nature of the data. In this case, intuition tells us that it makes sense to trade as soon as possible the next day as to maximize the use of the daily signal's alpha (or to minimize the time delay). Our optimization results indeed lead to such behavior. Intuition also tells us that high frequency signals should be useful to trade towards a daily or slowly moving target. In fact, we will write our optimization as a tracking problem which seeks to minimize the residual risk between our portfolio and the daily Markowitz portfolio.

One difference with most works is that we do not seek to find exact or numerical solutions to the HJB equations (except in some simple cases). Instead, using general properties of the solution we show how to construct simple approximations that can easily be used in practice. While we allow the presence of temporary market impact, we will be more interested in the case of linear costs. Hence, the focus will be on finding simple expressions for the boundaries of the NT or market-making zone. For the case of limit orders, we do not take into account partial execution. We find that in the presence of alpha predictors, one can be in a zone where it is optimal to send a market order. Furthermore, once we get closer to the NT zone (in this case a market-making zone), the use of a directional limit order is more desirable. In this paper we do not study in details the optimization over the limit order placement (i.e. its price). However, we point out how this can easily be included in our formalism. In order to test our algorithms, we use simple Monte Carlo simulations driven by mean-reverting alpha sources. We find that our strategies improve the global Sharpe of our P\&L compared to the case of simply trading using market orders to the daily Markowitz portfolio.

In the remaining of this section we present our notation, introducing few definitions that will be used in the following. In section 2 we describe the optimization problem with market orders only. We introduce the objective function and derive the relevant HJB equation. We propose an approximate solution of the HJB equation, and discuss the validity of the approximation. The trading strategy obtained from the approximate HJB solution is tested using a Monte Carlo simulation.  In section 3 we revisit the optimization problem, considering both limit and market orders. Also for this case, we derive an approximate solution of the relevant HJB equation and test the associated trading strategy using a Monte Carlo simulation. Appendix A contains basic facts about the  Ornstein-Uhlenbeck process, which is used to model the  predictors. Appendix B describes the HJB equation for the case of a general transient impact function. Appendix C, D and E, consider different regimes where the optimization problem simplifies. In particular, in appendix C we show that in the case of a zero volatility signal, the optimization problem can be solved exactly using the Euler-Lagrange variational  principle. In appendix D the HJB equation is solved exactly for the case of quadratic impact and zero linear costs.  In appendix E we focus on the case of large price impact and define a perturbative  scheme to solve the HJB equation recursively.

\subsection{Notation}
\label{notation}
We are allowed to trade from today's market opening time $t_\text{open}$ until today's market closing time $T$. The time interval between the close of subsequent market days is equal to $T$,  therefore yesterday's closing time  is  $0$ and tomorrow's closing time is $2T$.    We will optimize the integrated P\&L from some time $t$ included in today's trading hours, that is $t \in [t_\text{open},T]$,  until tomorrow's closing time, that is $2T$.  The relevant times are represented on the time's arrow in figure \ref{fig:time_axis}.

\myfig{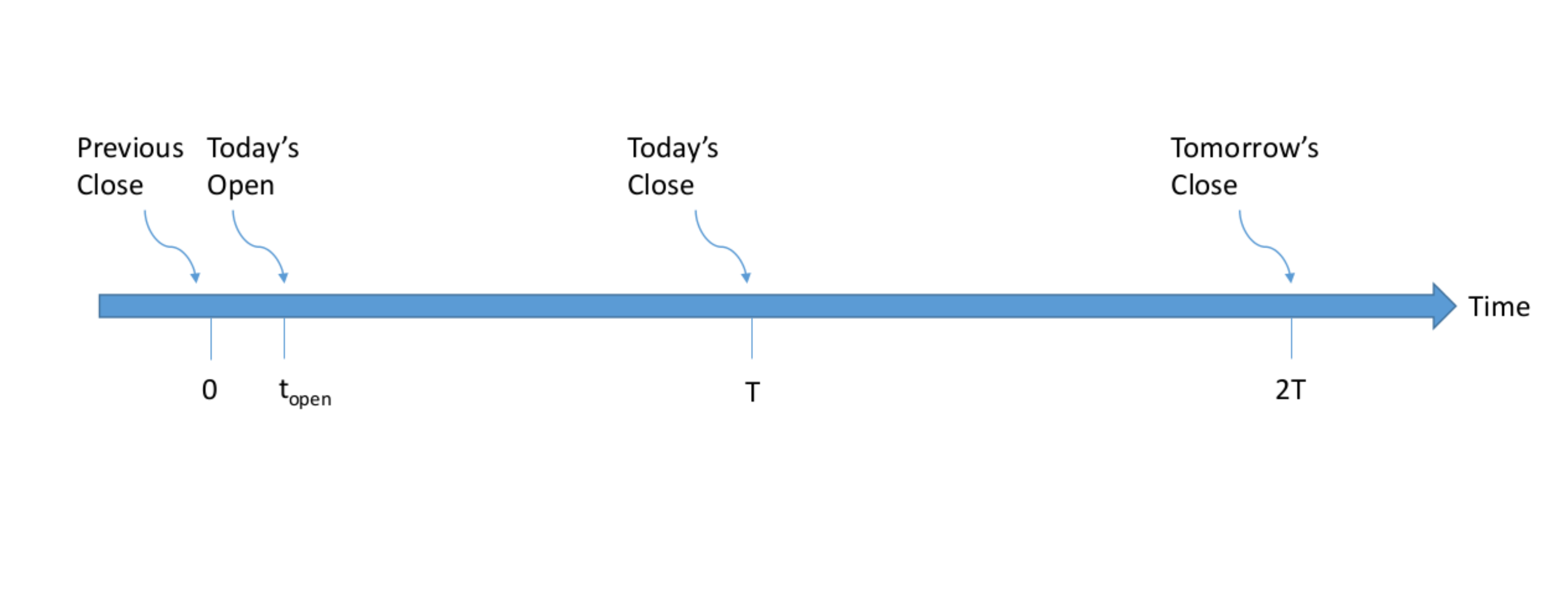}{0.4}{\small Trading time arrow. We denote today's opening time as $t_\text{open}$ and today's  closing time as $T$.  The previous day's close is at time $0$ and the next day's close is at time $2T$. }{time_axis}

For simplicity, we will consider a single asset with price $P_t$. The dynamics of the price is given by
\bea
dP_t = \alpha_t dt + \sqrt{\nu} dW_t
\eea
where $dt$ is the time scale of our trading decisions and $W_t$ is a standard Wiener process. The stochastic dependence in time will be denoted with a subscript, as in $dP_t$.    Without loss of generality, we will decompose the drift $\alpha_t $ into a constant $\bar \alpha$ and intraday component $x_t$ with zero mean:
\bea
\alpha_t := \bar \alpha + x_t\;,\;\;\;\; \mathbb{E}[x_t] = 0
\eea
where as usual,  $\mathbb{E}[\ldots]$ denotes  unconditional expectation and the conditional expectation is denoted as $\mathbb{E}[\ldots | \text{condition}]$. We can think of $\bar  \alpha$ as the alpha that comes from the daily predictors, while $x_t$ comes from the intraday or HF signals.   At this point we leave the dynamics of $x_t$ to be quite general:
\bea\label{gp} 
dx_t = \mu(t,x_t) dt + \sqrt{\eta(t,x_t)} dZ_t
\eea
where the drift $\mu(t,x_t)$ and variance $\eta(t,x_t)$ are allowed to depend on time $t$ and the signal $x_t$. We usually model the fast signal as a  Ornstein-Uhlenbeck process, that is a mean-reverting process with constant volatility:  $dx_t = -\kappa x_t dt + \sqrt{\eta }\,dZ_t$. Basics results concerning the Ornstein-Uhlenbeck mean-reverting process are collected in appendix \ref{mean}.

For later convenience, we  introduce the following differential operator:
\bea\label{hd}
\hat D_{t,x}:= \frac{\partial}{\partial t} + \mu(t,x)\frac{\partial}{\partial x} + \frac{1}{2} \eta(t,x) \frac{\partial^2}{\partial x^2} 
\eea
which is the infinitesimal generator of time translation and It\={o} diffusion for the process described in (\ref{gp}).  We define the integrated gain of the HF signal as:
\bea\label{gain}
g(t,x) := \int_t^{2T} \mathbb{E}\left[\left. x_s\right|x_t =x \right] ds
\eea
which encodes the gain associated to the intraday signal from time $t$ during today's trading hours to the end of tomorrow's trade. Using Feynman-Kac formula,  one can show that the  integrated gain obeys the following relation:
\bea\label{dg}
\hat D_{t,x} \cdot g(t,x) + x = 0
\eea

The evolution of our position $q_t$ is given by:
\bea\label{u}
dq_t = u_t dt
\eea
where $u_t$ is the rate of trading at time $t$. The  optimization problem boils down to finding the optimal rate of trading $u_t$ that minimizes a certain objective function. In general, we expect out trading rate to be a local function of time, our current position, and our predictor: $u_t := u(t,x_t,q_t)$.

In the following, we will make use of the positive part function $(x)_+$, defined  as  
\bea\label{pp}
(x)_+:=x\,\theta(x)
\eea  
with $\theta(x)$  is the Heaviside step function with $\theta(0)=0$.

\section{Optimization with market orders}
\label{mo}

In this section we study the case of trading using market orders. For simplicity of notation, we assume that the half-spread is constant. However, relaxing this restriction is quite simple  and our final (approximate) results are valid regardless of the dynamics of the spread. We focus on trading during a single day.  Therefore, we are allowed to trade from some time $t$ during open market hours, to the close of the present day, that is time $T$.  However, in order to take into account the overnight risk and possible alpha,  we will optimize the  integrated P\&L  from some time $t$ during the present day to the close of the \emph{next} trading day, that is $2T$.    

Under our assumptions, the relevant objective function is given by
\bea\label{omega}
\Omega(t,x,q) &=& \min_{\{u_s | s \in (t, T) \}} \mathbb{E}\left[\left. \int_t^T C |u_s| ds + K \int_t^T u_s^2 ds  \right.\right. \nonumber \\ 
&&\left.\left.-\int_t^{2T} \alpha_s q_s ds + \frac{1}{2} \lambda \nu \int_t^{2T} q_s^2 ds \right| q_t = q, x_t = x \right] 
\eea
where the optimal trading strategy is the one that gives the minimum  possible value for $\Omega(t,x,q)$.  The first term on the RHS of (\ref{omega}) is the cost of the market orders, where $C$ is   half of the bid-ask spread, which is assumed to be constant. The second term serves as a {\it control} or \emph{regulator} for the size of the trades. In particular, a large value for the constant $K$ implies small average trade size, and vice versa.  However, it can also be interpreted as coming from temporary market impact. In fact, we can also model a general impact function $ K \int |u_s|^p\, ds$ for a general exponent $p$. For simplicity, in the main text  we will focus on the quadratic impact function.  The case with a general exponent $p$ is discussed in appendix \ref{genp}.  Both the first and the second term in (\ref{omega}) encode contributions associated to the trading, therefore are integrated between the initial time  to the close of the present day,  given that we are optimizing the trading during a single day.

The third term is the gain coming from the alpha of the asset. This includes the contribution of the daily alpha $\bar \alpha$, which is constant during the trading day under consideration,   and the HF component $x_t$.    The last term in (\ref{omega}) is a risk control, and it is proportional to the variance of the P\&L. The constant $\lambda$ is a risk aversion parameter.  We notice that both the third and fourth term are integrated up to the close of the next trading day, in order to model the overnight risk and alpha.

Using the Hamilton-Jacobi-Bellman (HJB) principle, we can write the following relation for the objective function  $\Omega(t,x,q)$ defined in (\ref{omega}):
\bea\label{bel}
\Omega(t,x,q) &=& \min_{\{u_s | s \in (t, T) \}} \mathbb{E}\left[\left. (C |u|  + K  u^2   - (\bar\alpha + x) q  + \frac{1}{2} \lambda \nu  q^2) dt \right.\right. \nonumber \\ 
&&  + \Omega(t+dt,x+dx,q+dq) \bigg| q_t = q, x_t = x \bigg] 
\eea
Applying  It\={o}'s lemma, from relation (\ref{bel}) we can derive the following HJB equation for the objective function:
\bea\label{hjbo}
\hat D_{t,x}\cdot \Omega(t,x,q) - (\bar\alpha + x)q + \frac{1}{2} \lambda \nu q^2 + \min_u \left[ C|u| + K u^2 + \frac{\partial \Omega}{\partial q} u\right] = 0
\eea
where the differential operator  $\hat D_{t,x}$ is defined in (\ref{hd}).  In order to get a better intuition about the optimization problem, it is convenient to work with a slightly different version of the objective function (\ref{omega}). Indeed, as we will show,  we can write the optimization as a {\it portfolio tracking problem}. For that, we introduce the daily Markowitz portfolio:
\bea
\bar q := \frac{\bar \alpha}{\lambda \nu}
\eea
and a new objective function:
\bea\label{v}
V(t,x,q) := \Omega(t,x,q) + g(t,x) q  + \frac{1}{2} \lambda \nu \bar q^2 (2T- t)
\eea
where $g(t,x)$ is the gain associated to the HF signal, as defined in equation (\ref{gain}). The two functions  $V(t,x,q)$ and $\Omega(t,x,q)$ differ by terms that do not depend on the trading rate $u_s$, which implies that they are modeling the same optimization problem.  That is, we can use indistinctly one of the two objective functions to compute the optimal trading rate. In the following we will always work with  $V(t,x,q)$.

To understand  the meaning of the objective function $V$ it is useful to write it as
\bea\label{vex}
V(t,x,q) &=& \min_{\{u_s | s \in (t, T) \}} \mathbb{E}\left[\left. \int_t^T \left(C|u_s|- g(s,x_s) u_s\right) ds \right.\right. \nonumber \\ 
&&\left.\left. + K \int_t^T u_s^2 ds +\frac{1}{2} \lambda \nu \int_t^{2T} (q_s-\bar q)^2 ds \right| q_t = q, x_t = x \right] 
\eea
The first term is the effective cost of the market order,  once we take into account the HF predictor. It follows that the intraday signal can be used to reduce the cost of trading, which in principle can also become negative.   The last term is the integrated residual risk (variance) of the difference between our position and the daily Markowitz portfolio. This is the term that pushes the trading toward the daily Markowitz portfolio $\bar q$, which is the target position suggested by the daily signal.  It follows that the original optimization problem is dual to tracking the Markowitz portfolio using the fast signals.  This decomposition, while rather arbitrary, is very useful in practice as the trader can simply replace $\bar q$ with the ideal portfolio which comes from a daily back-test.

Using the definition (\ref{v}), and the HJB equation for $\Omega$ (\ref{hjbo}), it is possible to write down an HJB equation for V:
\bea\label{hjbv}
\hat D_{t,x}\cdot V+ \frac{1}{2} \lambda \nu (q-\bar q)^2 + \min_u \left[ C|u| + K u^2 + \left(\frac{\partial V}{\partial q} - g \right) u\right] = 0
\eea
where we used the relation (\ref{dg}). The  boundary condition follows from the expression (\ref{vex}) and it is given by:
\bea\label{bcv}
V(T,x,q) = \frac{1}{2} \lambda \nu T (q-\bar q)^2
\eea
which  is precisely the residual variance around the Markowitz portfolio, at the end of the day.   The optimization problem boils down to solving the HJB equation (\ref{hjbv}) with boundary condition (\ref{bcv}).

In order to reduce the equation (\ref{hjbv}) to a standard PDE, we  derive the trading rate $u$ that minimizes the term $\left[ C|u| + K u^2 + \left(\frac{\partial V}{\partial q} - g \right) u\right]$. In this way, the trading rate $u$ is related to  the gain $g$, the linear costs $C$, and  the derivative of the objective function respect to the initial position, that is $\frac{\partial V}{\partial q}$. The  explicit expression for  $u$ depends on the location in the $(t,x,q)$ space. In particular,  the $(t,x,q)$ space is divided into three regions and in each of these regions $u$ assumes a different expression\footnote{Another possible formulation of the problem is to assume constant and discretized   trading rate, that is  $u \in [-Q, 0,  Q]$. We will consider this model in the next section, where we allow also limit orders.}. These are the regions where we buy, sell or don't trade.   In details, the three  zones and the corresponding trading rate $u$ are given by:
\begin{enumerate}
\item $g>C+\frac{\partial V}{\partial q}$. In this case $u>0$, so we \emph{buy}:
\bea\label{buy}
u =\frac{1}{2 K } \left(g-C- \frac{\partial V}{\partial q}\right) 
\eea
\item  $g<-C+\frac{\partial V}{\partial q}$. In this case $u<0$, so  we \emph{sell}: 
\bea\label{sell}
u =-\frac{1}{2 K } \left(-g-C+ \frac{\partial V}{\partial q}\right)
\eea
\item  $-C+\frac{\partial V}{\partial q}\le g \le C+\frac{\partial V}{\partial q}$.  In this case we \emph{ don't trade}, that is:
\bea
u = 0
\eea
\end{enumerate}
The boundary between the sell zone  and the no-trading (NT) zone is a two dimensional manifold embedded in the  $(t,x,q)$ space. We describe it via an explicit parameterization $q=b_-(t,x)$. In the same way, the boundary between the buy zone and the NT zone is described as  $q=b_+(t,x)$.  From equations (\ref{buy}),(\ref{sell}), it follows that $b_\pm(t,x)$ are implicitly defined by the following equations:
\bea\label{beq}
g\mp C=\left.\frac{\partial V}{\partial q}\right|_{q=b_{\pm}}
\eea
An optimal trading trajectory that starts from the buy zone, enters the NT zone and stays there until the end  of the day  is sketched in  figure \ref{fig:zones}. As we show in appendix \ref{detsig}, this scenario is realized for the case of a  mean reverting signal with zero volatility. For the case of a stochastic signal, in general one can go in and out of the NT region many times during the day. As we can see in figure \ref{fig:zones}, the width of the NT zone increases towards the end of the day. This means that we are more aggressive at the beginning of the day. Note that this is consistent with our intuition under the presence of a daily signal. Since the daily signal is simply a drift, we maximize our profit by trading at the beginning of the day.
\myfig{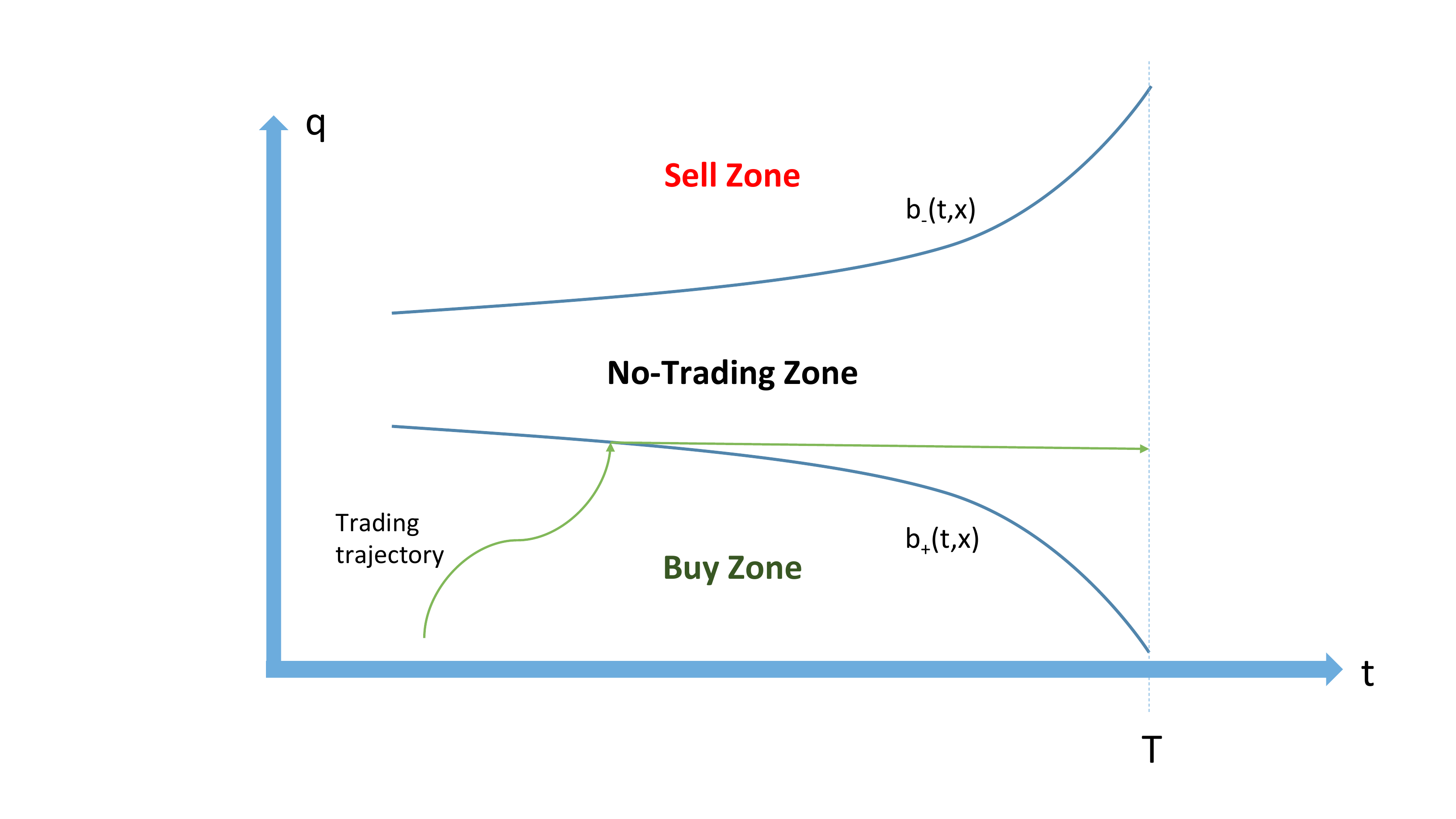}{0.4}{\small  The sell, buy and no-trading  (NT)  zones are shown in this figure. We denote as $b_-(t,x)$ ($b_+(t,x)$) the boundary between the sell (buy) region and the NT region. We show a trading trajectory with initial position in the buy zone. The trading moves the position toward the NT zone, and when this is reached, the trading stops. }{zones}

We can summarize the trading rate $u$ as:
\bea\label{um}
u= \frac{1}{2 K } \left(g-C- \frac{\partial V}{\partial q}\right)_+-\frac{1}{2 K }  \left(-g-C +\frac{\partial V}{\partial q}\right)_+ 
\eea
where the positive part  function  $(x)_+$  is defined in (\ref{pp}).  The equation for $V$ (\ref{hjbv}) can now be written in compact form:
\bea\label{hjbm}
\hat D_{t,x} \cdot V + \frac{1}{2} \lambda \nu (q - \bar q)^2 - K u^2 = 0
\eea
where $u$ is given in (\ref{um}) and the boundary condition is defined in (\ref{bcv}). The HJB equation (\ref{hjbm}) is very difficult to solve in general.   In the remaining of this section we consider  the approximation where the source term $-K u^2$ in equation (\ref{hjbm})  is neglected. We argue that this is a good approximation for a market without price impact, that is  in the limit $K\rightarrow 0$,  and describe the implementation of our trading strategy for this scenario using a Monte Carlo simulation.

In the case where the signal has zero volatility, there is no uncertainty and the optimization problem can be solved using the Euler-Lagrange variational principle. We discuss this case in appendix \ref{detsig}, where we are able to derive an explicit expression for the boundaries $b_\pm$ for any deterministic signal. We also compute the optimal trading trajectory for the case of a deterministic mean revering signal, showing that for this case, once we enter the no-trading (NT) zone, we stay there until the end of the day.  Another general feature of the solution is that as $K\rightarrow 0$ we go instantaneously to one of the boundaries of the NT zone (if we are out).

Another exactly solvable case is when we set the linear costs to zero, that is  $C = 0$. In this case, we are left with  the quadratic costs and the problem becomes very similar to the discrete-time optimization studied in \cite{garlenau}. In appendix \ref{quadcost} we solve this problem explicitly for a general signal using the Feynman-Kac theorem.

We notice that the problem discussed in  \cite{martin} essentially   corresponds to the optimization problem described in this section, but in the case of an infinite time horizon  and  in the regime where the boundary of the NT region is reached instantaneously (no price impact).

\subsection{Approximate HJB solution}
\label{moapprox}

We have shown how the optimal strategy can be expressed in terms of the objective function,  which is defined by the HJB  equation (\ref{hjbm}).  An exact analytical solution of this equation  seems out of reach, therefore for practical purposes, we propose using the solution that comes from ignoring the source term $-K u^2$ in (\ref{hjbm}), that is:
\bea\label{apx}
V \approx \frac{1}{2} \lambda \nu (2T-t) (q- \bar q)^2
\eea
Using this approximate objective function $V$,  we obtain
\bea
\frac{\partial V}{\partial q} &=& \lambda \nu (2T -t) (q - \bar q)
\eea
and equation (\ref{beq}) can be solved explicitly to obtain the boundaries $b_\pm(t,x)$.  It results:
\bea\label{bexp}
b_\pm(t,x) = \bar q + \frac{1}{\lambda \nu (2T - t)} \left( g(t,x) \mp C\right)
\eea
We notice that (\ref{bexp}) is equivalent to the correspondent  expression for the deterministic case (\ref{detb}),  where the deterministic gain is replaced by the stochastic gain.

In which regime can we justify the approximation (\ref{apx})?
In the limit $K \rightarrow \infty$, the trading rate goes to zero as $u \sim 1/K$, therefore also   the source term goes to zero as $-K u^2\sim 1/K$.   It follows that in this regime, as a first order approximation we can neglect the source term and work with the solution in (\ref{apx}).  In fact, one can also define a systematic expansion in $1/K$, as we describe in  appendix \ref{kinf}.

The other, more practical limit is when $K \rightarrow 0$, that corresponds to the scenario where the price impact is negligible and  is a good approximation  for the case of  a small trader.     In this case, as suggested by intuition and the results for the deterministic case in appendix \ref{detsig},   we trade towards the boundary of the NT zone instantaneously (when we are out of the NT region). One can argue that in this case $u$ diverges, but in practice, the trade size must be finite, that is 
\bea
dq = u\, dt  = \text{finite}
\eea
Moreover, in practice $dt$ is finite (e.g. one second), therefore $u$ remains finite as $K$ goes to zero and we can justify our approximation $-K u^2\approx 0$. This is, of course, not a formal mathematical proof but only an argument that will allow us to use the result (\ref{bexp}) for the case when the market impact can be neglected.

\subsection{Implementation and simulation}
We now describe how our formalism can be implemented for real trading. We consider the case of a small trader, where  the price impact can be ignored.  As we have already discussed, this corresponds to the limit $K\rightarrow 0$, where the approximation (\ref{apx}) is reliable. In this regime, we trade instantaneously towards one of the boundaries if we are out of the NT zone. In particular, we trade to the $b_-(t,x)$ boundary if we are in the sell zone and to the boundary $b_+(t,x)$ if we are in the buy zone. The expression for the boundaries $b_\pm(t,x)$ is given in (\ref{bexp}).    The basic trading decision scale is set to $dt$.

We can take the daily Markowitz positions $\bar q$ to be the ideal positions that come from a daily back test.  Therefore, the Lagrange multiplier $\lambda$ can be written as:
\bea
\lambda \approx \frac{\text{Annualized Sharpe Ratio of Daily Target}}{\text{Annualized Volatility of Daily Target}}
\eea

We assume that we can model high-frequency predictors as mean-reverting processes.  In practice, it is useful to decompose our intraday alpha $x_t$ in terms of a z-score or signal $\epsilon_t$ with $\mathbb{E}[\epsilon_t] = 0$ and $\text{Var}[\epsilon_t] = 1$.  We can write:
\bea\label{beta}\nonumber
x_t &:=& \beta \sqrt{\nu} \epsilon_t  \\
d\epsilon_t &=& -\kappa \epsilon_t dt + \sqrt{2\kappa} dZ_t 
\eea

The constant $\beta$ can be conveniently written in terms of the annual Sharpe ratio of the ideal HF position $\tilde q_t = \epsilon_t/\sqrt{\nu}$:
\bea
\beta \approx \frac{\text{Annualized Sharpe Ratio of HF Signal}}{\sqrt{252 T}}
\eea
 The gain is then:
\bea
g(t,\epsilon) = \frac{\beta \sqrt{\nu}\epsilon}{\kappa} \left(1 - e^{-\kappa(2T-t)}\right) \approx \frac{\beta \sqrt{\nu} \epsilon}{\kappa} 
\eea
where the last approximation is valid if the time scale of the predictor is much smaller than one day.

We have tested our algorithm with a Monte Carlo simulation, the results are shown  in figure \ref{fig:market}.  In the simulation, the decision time scale is $dt = \text{1 minute}$ and when we are out of the NT zone we trade directly to  the  boundary of the NT zone in one shot.  The intraday signal has a mean-reversion of $30$ mins, and the daily signal is constant during the day, but varies from day to day with a mean-reversion time scale of $10$ days and a annual Sharpe of $2.1$. Other relevant parameters are the price variance $\nu = 0.01$, the half spread $C = 0.01$,  the risk aversion  $\lambda = 37.4$. We parameterize the HF as in (\ref{beta})  with  $\beta = 1$,  which corresponds to a HF Sharpe = 16.5.

In figure \ref{fig:market}, the blue line is the cumulative P\&L for the daily signal, where the trading is all at the beginning of the day, and the linear costs are ignored. The green line is the same P\&L with the linear costs taken into account. The red line is the P\&L for our strategy obtained from the HJB approach, which takes into account also the trading costs. We see that, thanks to the HF signal and the intraday trading, our strategy outperforms the daily strategy. Indeed, our P\&L is higher than the daily P\&L, also  when the trading costs are ignored,  which  means that thanks to the HF signal we can follow the daily signal with negative trading costs.

\myfig{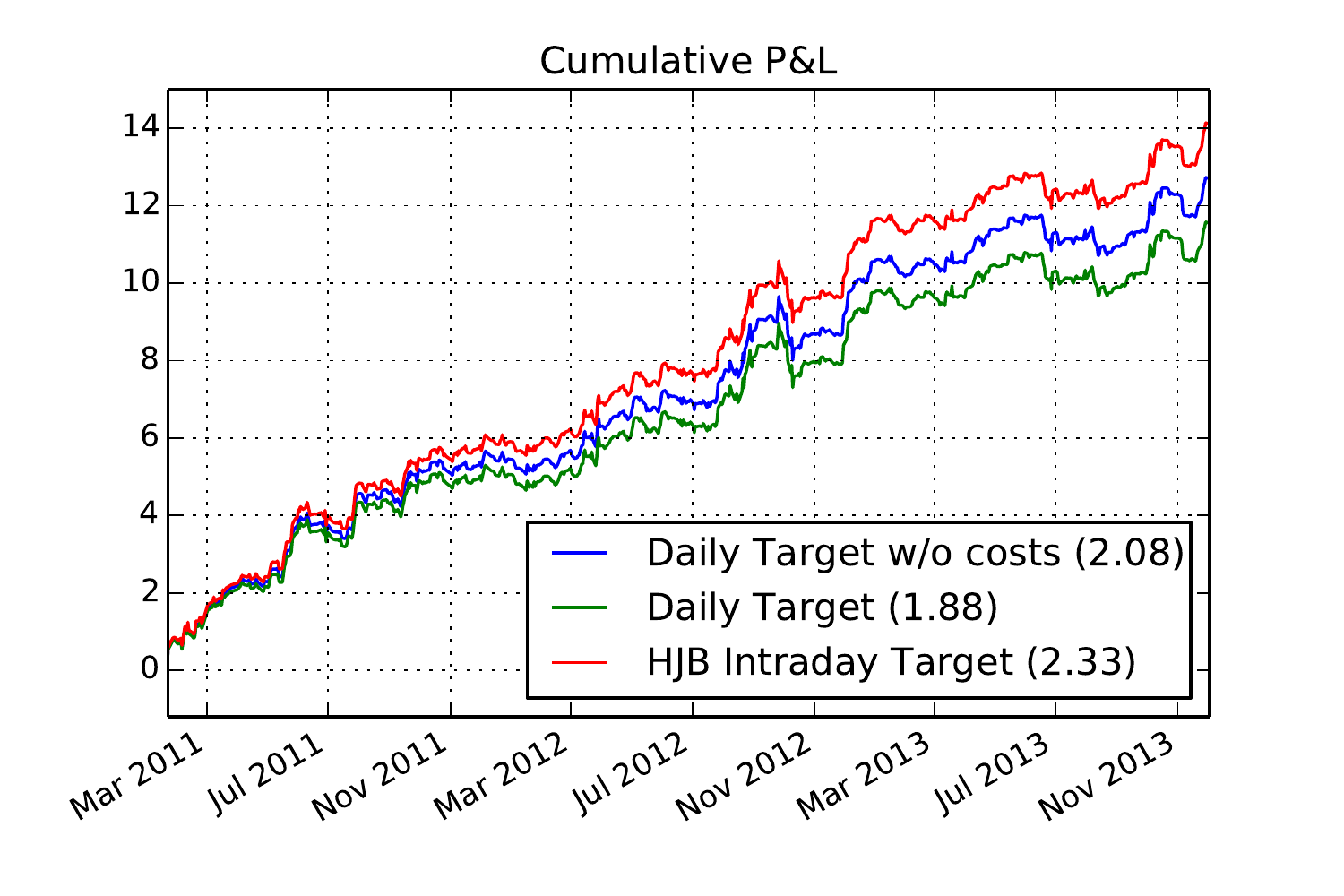}{.8}{\small This figure shows several cumulative P\&L's, where the numbers in parenthesis in the legend are the correspondent Sharpe ratios.  The blue line  is the  P\&L of a daily strategy without taking into account the linear costs. The green line is the same strategy with the linear costs taken into account. The red line is the HJB strategy with intraday trading, that also takes into account the trading costs. The HF mean-reversion scale is  $30$ mins and the  daily signal mean-reversion time scale is $10$ days.  The relevant parameters are:    $\nu = 0.01$,  $C = 0.01$,   $\lambda = 37.4$,  $\beta = 1$.
}{market}

\section{Optimization with limit orders}
We now revisit the optimization problem described in the previous sections,  allowing also limit orders.  We can use the same HJB framework, where now the trading rate $u_t$ defined in (\ref{u})  can get a contribution from  market orders and limit orders.  In details,  our position will now evolve according to:
\bea
dq_t= (m_t^+ - m^-_t  + \mathbf{1}_t^+ l^+_t  - \mathbf{1}_t^- l^-_t) dt\;,\;\;\;\; m^\pm_t,l_t^\pm \geq 0 
\eea
where $m_t^\pm$ and $l_t^\pm$ are the sizes of the market and limit orders respectively. The indicator $\mathbf{1}_t^\pm$ equals one if the limit order is executed in the interval $dt$ and zero otherwise. Hence, we assume that our limit orders are small enough to be filled in one shot (we ignore partial fills).

For simplicity, we will assume that the spread is constant (one tick) and hence limit orders are only executed if there is a mid price move in the right direction. As in the previous section, we denote as $C$ the half spread.  At the end of the interval $dt$ all pending limit orders are canceled.
We also assume that we place limit orders at the top of the book and hence do not optimize for their price. We will comment on how to implement a price optimization at the end of the section.

Let $P^+,P^-$ be the conditional probability that a limit buy and sell order will be filled within the next time step $dt$. 
So far we have modeled the price in our decision interval $dt$ as a diffusion process. However, in real life, there are price movements inside $dt$ which are discrete (tick-by-tick). These are the movements which must be predicted by $P^\pm$.  Hence for limit orders, we must have a very short term predictor $y_t$ which might be different from the longer-term intraday alpha $x_t$. Hence, the fill probabilities can be a function of time and our different intraday alpha streams:
\bea
P^\pm(t,x,y): = \mathbb{E}[\mathbf{1}_t^\pm|x_t = x,y_t = y]
\eea
In practice, we expect $y_t$ to dominate the fill probability by having a much shorter mean-reversion time scale than $x_t$. Moreover, ideally we would have both alpha sources being uncorrelated. We will come back to this point in section \ref{impsim}.

We will ignore price impact and simply cap the size of our trades so that:
\bea
m^\pm &\in& [0,Q]\\
l^\pm &\in& [0,Q] \\
m^\pm + l^\pm &\in& [0,Q]
\eea
where $Q > 0$. Note that, as will become apparent below, without the impact term the last restriction will simply pick either the limit or market order.

The HJB equation (\ref{hjbv}) takes the following form:
\bea\label{hjbl}
0 &=& \hat D_{t,x,y}\cdot V+ \frac{1}{2} \lambda \nu (q-\bar q)^2 \nonumber \\
&&+ \min_{m^\pm,l^\pm} \left[ m^+\left( C + \frac{\partial V}{\partial q} - g \right)
- P^+ l^+ \left(C  - \frac{\partial V}{\partial q}  + g\right) \right. \nonumber \\
&&\left. + m^-\left( C - \frac{\partial V}{\partial q} + g \right)
 - P^- l^- \left(C  + \frac{\partial V}{\partial q}  - g\right) \right] 
\eea
where, as in the previous section,  $C$ is half of the bid-ask spread and it is assumed to be constant, and $\hat D_{t,x,y}$ is the infinitesimal generator of time and It\={o} diffusion for our set of predictors.

Our trading decisions are based on the optimization over $m^\pm,l^\pm$ in the HJB equation (\ref{hjbl}). As in the case of market orders, we find different trading regions.  In particular, we find a  sell region,  a buy region and a market-making region. Both the sell and the buy regions are divided into two subregions,  that is the market orders region, where we send market orders, and the limit orders region, where we send  limit orders. In total, our strategy includes five regions: a region where we send sell market  orders, a region where we send sell limit  orders, a region where we do market-making, a region where we send  buy limit orders and a region where we send buy  market orders.

In details, the five regions we just described, are defined by the following expressions:
\begin{enumerate}
\item  $g>C\frac{1+P^+}{1-P^+}+\frac{\partial V}{\partial q}$  we send a \emph{buy market}  order.
\item  $C+\frac{\partial V}{\partial q}<g<C\frac{1+P^+}{1-P^+}+\frac{\partial V}{\partial q}$  we send a \emph{buy limit}  order.
\item  $-C+\frac{\partial V}{\partial q}\le g \le C+\frac{\partial V}{\partial q}$ we \emph{send both a buy and a sell limit order} (market-making).
\item  $-C\frac{1+P^-}{1-P^-}+\frac{\partial V}{\partial q} <g<  -C+\frac{\partial V}{\partial q}$  we send a \emph{sell limit} order.
\item  $g< -C\frac{1+P^-}{1-P^-}+\frac{\partial V}{\partial q}$  we send a \emph{sell market} order.
\end{enumerate}
We note that the boundaries of the market-making region are the same as the NT zone in the case of  market orders only. Therefore, using the same notation as in the previous section, they are defined by 
\bea\label{ib}
g\mp C=\left.\frac{\partial V}{\partial q}\right|_{q=b_{\pm}}
\eea
where $b_+(t,x)$ ($b_-(t,x)$) now separates the market-making region from the  buy  (sell) limit region. We define  $\tilde b_+(t,x, P^+)$  as the boundary between the buy market region and the buy limit region  and $\tilde b_-(t,x, P^-)$ as the boundary  between the sell market region and the sell limit region. They are implicitly defined by the following equation 
\bea\label{tib}
g\mp C\frac{1+P^\pm}{1-P^\pm}=\left.\frac{\partial V}{\partial q}\right|_{q=\tilde{b}_{\pm}}
\eea
The five trading regions and all the boundaries are sketched in figure \ref{fig:zones2}. This is a schematic representation, where $x,P^+,P^-$ are constant during the day. In reality, they are stochastic quantities.  

Note that the strategy requires market orders when we are far from the  market-making region (that is we are far from the ideal daily position $\bar q$), and requires limit orders when we are close to it. Therefore, the strategy becomes more aggressive when we are far from the optimal position, in agreement with intuition. 

\myfig{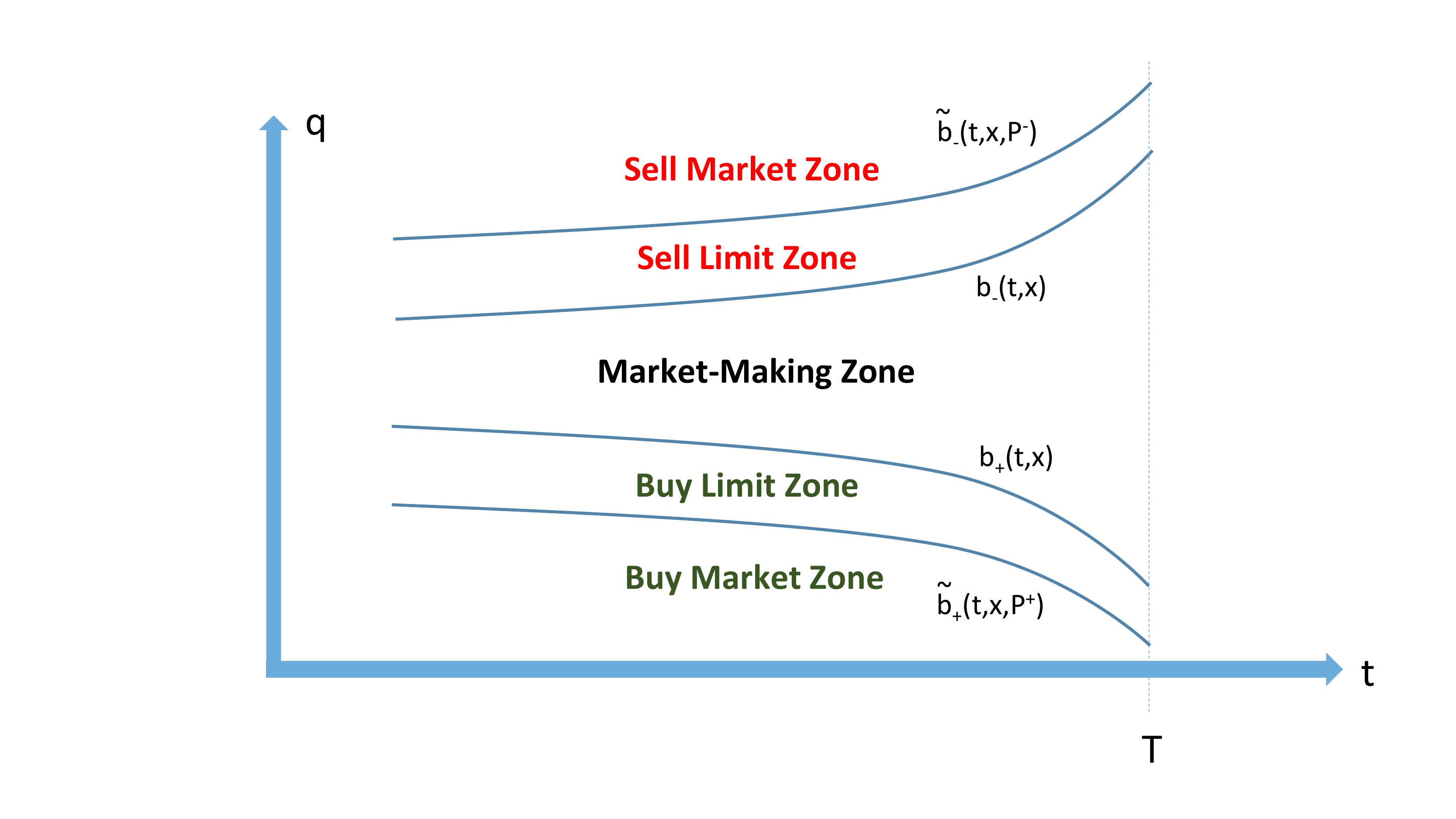}{0.4}{\small  The five zones for the case with limit orders are shown in this figure. We denote as $b_-(t,x)$ ($b_+(t,x)$) the boundary between the  sell (buy) limit region and the market-making region. The boundaries between the  sell (buy) market region and the sell (buy) limit region are denoted as $\tilde b_-(t,x, P^+)$ ($\tilde b_+(t,x, P^-)$). For graphical purposes, we represent the zones for the case of  constant $x,P^+,P^-$.}{zones2}

In section \ref{impsim}, we test our strategy  using a Monte Carlo simulation. In our simplified model  we find that it is suboptimal to send both buy and sell limit orders at the same time (market-making). The reason is that, in our simple simulation,  we never get both orders executed at the same time, and we are forced to trade more in the future.  Instead, it is optimal to not trade in this region.  Therefore,  in our simulations we will replace the market making region with a no-trading region.

What is the optimal values for the trading rate $Q$? 
As in the case of market orders, in the absence of impact, we would like to trade as much as possible. However, we should not cross trading regions because this will lead to more unnecessary back-and-forth trading in the future. 
Hence, we argue that it is optimal to trade to the boundary of our current trading region.

Note that we have not optimized for the price of the limit orders as in \cite{lehalle1,lehalle}. Even though we will not study this problem in detail in this paper, we  should mention that this optimization is simple to implement in our framework. For example, assuming that the spread is always one tick, let $C+\delta^\pm$ be the price distance of the limit orders from the mid price, where $\delta^\pm \geq 0$.  The fill probability for a limit order with a price distance $C+\delta^\pm$, is defined as
\bea
P^\pm(\delta^\pm,t,x,y): = \mathbb{E}[\mathbf{1}_{\delta^\pm, t}^\pm|x_t = x,y_t = y]
\eea
Then our optimization problem now includes two more dimensions:
\bea
&&\min_{m^\pm,l^\pm,\delta^\pm} \left[ m^+\left( C + \frac{\partial V}{\partial q} - g \right)
- P^+(\delta^+,t,x,y)\,l^+\left(C + \delta^+  - \frac{\partial V}{\partial q}  + g\right) \right. \nonumber \\
&&\left. + m^-\left( C - \frac{\partial V}{\partial q} + g \right)
 - P^-(\delta^-,t,x,y)\, l^-\left(C + \delta^-  + \frac{\partial V}{\partial q}  - g\right) \right] 
\eea
The detailed study and simulation of the problem including optimal order placement will be left for future work.

\subsection{Approximate HJB solution}
\label{ap}
As in the case of market orders, a full solution of the HJB  equation seems hopeless.
Therefore in this case, we also approximate $V$ to be the value function of the no-trading zone:
\bea 
V(t,x,q) \approx \frac{1}{2} \lambda \nu (2T -t) (q - \bar q)^2
\eea
which gives
\bea
\frac{\partial V}{\partial q} &=& \lambda \nu (2T -t) (q - \bar q)
\eea
Equations (\ref{ib}) and (\ref{tib}) can be solved explicitly to obtain the boundaries $b_\pm(t,x)$ and $\tilde b_\pm(t,x, P^\pm)$.  It results:
\bea\nonumber\label{exptb}
b_\pm(t,x) &=& \bar q + \frac{1}{\lambda \nu (2T - t)} \left( g(t,x) \mp C\right)\\
\tilde b_\pm(t,x, P^\pm) &=& \bar q + \frac{1}{\lambda \nu (2T - t)} \left( g(t,x) \mp C\frac{1+P^\pm}{1-P^\pm}\right)
\eea

\subsection{Implementation and simulation}
\label{impsim}
We have tested our trading strategy using a Monte Carlo simulation. In particular, we work in the approximation described in  section \ref{ap}, where he boundaries of the various regions are given in equations (\ref{exptb}). As we already mentioned, we are in the approximation of zero impact factor, therefore we trade instantaneously to the boundary of the region in which we are. 

For our simulations, in order to capture the very short-term behavior of the price relevant to limit orders, we decompose our intraday alpha as a \emph{slow} ($\epsilon_t$) and \emph{fast} ($\tilde \epsilon_t$) predictors. That is:
\bea
dP_t &=& (\bar \alpha + x_t)dt + \sqrt{\nu} dW_t \nonumber\\
x_t &=& \sqrt{\nu} \left( \beta \epsilon_t +  \tilde \beta \tilde \epsilon_t\right) \nonumber \\
d\epsilon_t &=& -\kappa \epsilon_t dt + \sqrt{2\kappa}\,dZ_t \nonumber \\
d\tilde \epsilon_t &=& -\tilde \kappa \tilde \epsilon_t dt + \sqrt{2\tilde \kappa}\, d\tilde Z_t
\eea 
The fast signal will dominate the short term price predictability only if:
\bea
 \tilde \beta \gg \beta\;,\;\;\;\; \frac{\tilde \beta}{\tilde \kappa} \ll \frac{\beta}{\kappa}
\eea
In fact, the mean reversion scale of $\tilde \epsilon_t$ will be of the order of $dt$:
\bea
\tilde \kappa \sim 1/dt
\eea
In this case, the fill probabilities will be functions of $\tilde \epsilon$ only, while the long-term integrated gain will be only a function of $\epsilon_t$:
\bea
P^\pm \approx P^\pm(\tilde \epsilon)=\Phi\left(\frac{\mp \sqrt{\nu}\tilde \epsilon-\frac{2 C}{dt}}{\sqrt{\frac{\nu}{dt}}}\right)\;,\;\;\; g(t,\epsilon) \approx \frac{\beta \sqrt{\nu} \epsilon}{\kappa}
\eea
where $\Phi$ is the cumulative function of the unit normal distribution. In the figure \ref{fig:limit} we show the result of a Monte Carlo simulation with our simple limit/market order algorithm.
All parameters are the same as in our simulation in section \ref{mo}. For the fast  signal $\tilde \epsilon_t$, we take a  mean-reversion time scale of $1$ minute (that is, it is equal to $dt$) and $\tilde \beta = 13$.  As in the previous simulation, we trade instantaneously towards the boundary of our zone.  Even though our simulation does not take into account the market making, we find a mild improvement from the use of limit orders.

\myfig{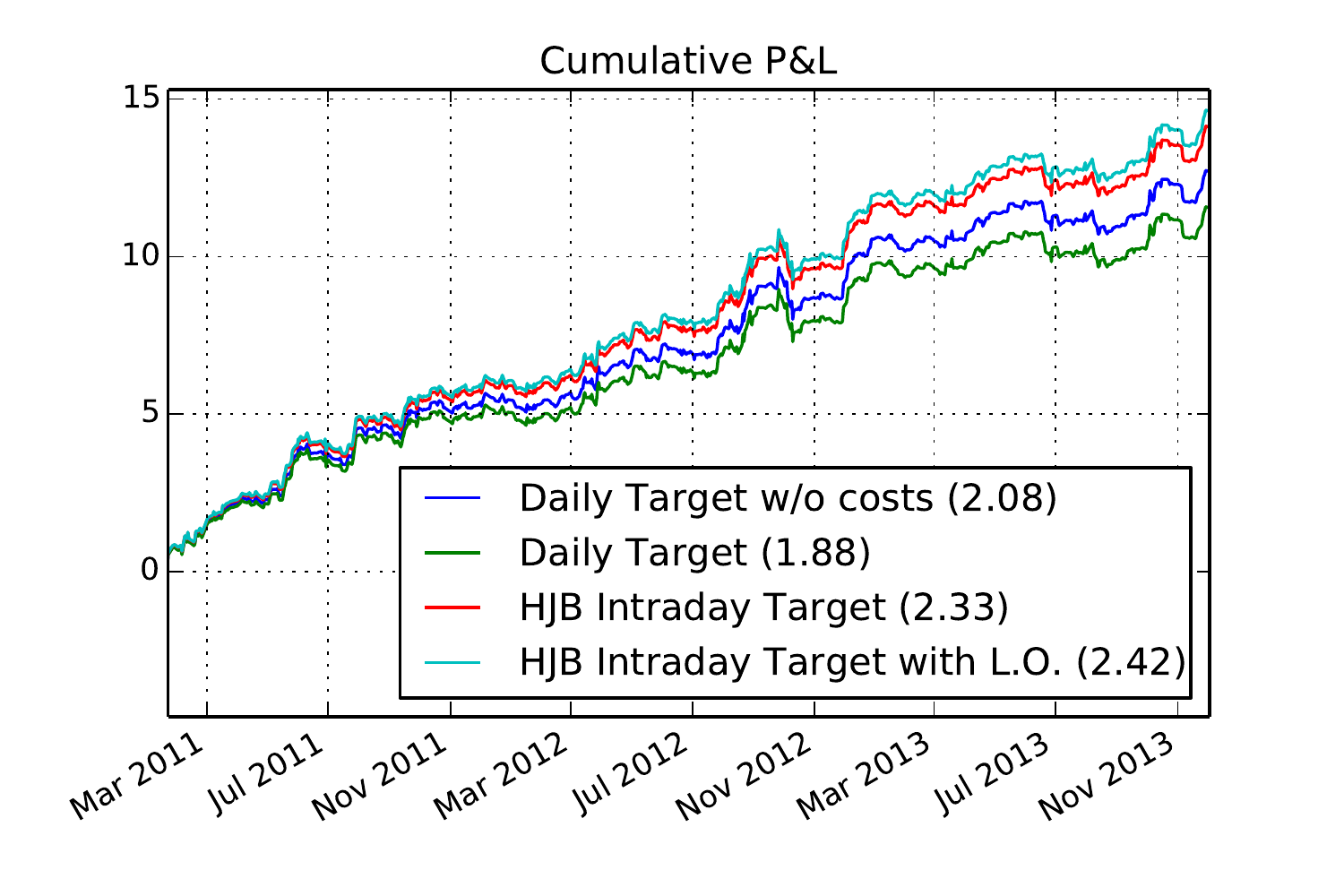}{0.8}{\small The first three cumulative P\&L in the legend are the same trading strategy described in figure \ref{fig:market}. The last P\&L  in the legend (the light blue line), is associated to the HJB strategy with intraday trading, with market and limit orders. Trading costs are also taken into account. The relevant parameters are as in figure  \ref{fig:market} and the fast predictor has  mean-reversion time scale of $1$ minute and $\tilde \beta = 13$.
}{limit}

\section{Conclusions}
\label{sec:conclusion}

In this paper  we have developed a general framework to derive optimal trading strategies which exploit any possible source of  alpha.  We used the  Hamilton-Jacobi-Bellman (HJB) theory,  and  even though the HJB equations cannot be solved exactly (except in few simplifying limits),  we have presented several algorithms which are inspired on general features of the exact solution and can be easily implemented in real trading. 

Our framework uses finite horizon optimization. Indeed, we optimize the trade during one single day and we  take into account the overnight risk and possible overnight alpha.  We  isolate the daily component of the predictors from the intraday one.  In this way, the  daily alpha that is constant during any single trading day, defines an ideal position or Markowitz portfolio, which the trader should reach. Our strategy  specifies how the intraday signals can be exploited to reach the ideal position dictated by the daily signal.

We take into account two sources of slippage:  temporary market impact and linear costs. The former implies a reduction of  the trading rate.  The latter, in the case where we  allow only market orders, gives rise to a no-trading (NT) region centered around the Markowitz portfolio and with boundaries that depend on the value of the intraday predictors.  A trader with a position outside of the  NT region trades  toward the boundary of the NT region.   See figure  \ref{fig:zones}.

In the case where we allow both market and limit orders, the NT region is replaced by a market-making region, where the trader sends a buy and a sell limit order together. In this set up, the buy and the sell regions are divided each one into two subregions, characterized by the type of orders that are used. For instance, the sell region is divided into a sell limit region,  where the trader sends   sell limit orders, and a sell market region where the trader sends  sell market orders.  The market order regions are the most distant from the  Markowitz portfolio,    therefore in agreement with intuition,  in these regions we trade  more aggressively sending market orders.  See figure  \ref{fig:zones2}.

To summarize, our framework combines limit and market orders in a unifying way and  allows the trader to exploit daily alpha signals, intraday alpha signals and  market-making around the optimal daily position.   

We have tested our analytical algorithms using Monte Carlo simulations, focusing on the case of a small trader where the price impact can be neglected.  The numerics confirm that our strategies increase the cumulative Profit and Losses respect  to a trader which naively follows the Markowitz portfolio.  

In this paper we have not optimized for the price of the limit orders, but simply considered the orders which are the closest to the mid price (that is, limit orders at  the top of the order book).  Furthermore, our simple Monte Carlo  did not allow for a numerical simulation of the market-making, and we replaced this region with a no-trading region. It is natural to expect that a proper optimization of the limit orders price and the inclusion of market-making will further improve the Profit and Losses in our simulations.    We leave this for future studies.

\section*{Acknowledgments}
 We would like to thank Adela Baho, Slava Belyaev, Arthur Berd,  Jim Gatheral,  Petter Kolm, Jung-Woo Lee and  Matthew Lorig for discussions.  The work of  F.P. is supported by a Marie Curie International Outgoing Fellowship  FP7-PEOPLE-2011-IOF, Project n\textsuperscript{o} 298073 (ERGTB).

\appendix

\section{Ornstein-Uhlenbeck mean reverting  process}\label{mean}

The Ornstein-Uhlenbeck process is described by:
\bea\label{uo}
d x_t=\kappa(\bar x -x_t) dt +\sqrt{\eta} d Z_t
\eea
which implies that the process reverts around the value  $\bar x$ with a reversion rate equal to $\kappa$. In the main text we usually consider $\bar x=0$,   however the results in this appendix are derived for a general $\bar x$ for completeness. The variance $\eta$ is taken to be constant and $Z_t$ is a standard Wiener process. The solution of the  (\ref{uo}), with boundary condition  $x_t=x$, is given by     
\bea
x_{s}=\bar x +(x-\bar x) e^{-\kappa (s-t)}+\sqrt{\eta} \int_t^{s} e^{-\kappa (s-r)} d Z_r
\eea
$x_s$ is normally distributed and  its expectation and variance, conditioned by the boundary condition $x_t=x$, are   given by 
\bea\nn\label{stat}
\mathbb{E}\,\left[ x_s|x_t=x\right]&=&\bar x +(x-\bar x) e^{-\kappa (s-t)}\\
\text{Var}\,\left[ x_s|x_t=x\right]&=&\frac{\eta}{2\kappa}(1-e^{-2\kappa(s-t)})
\eea
Considering the process $x_s$ as an HF signal, we can compute its gain $g(t,x)$ as defined in formula (\ref{gain}). It results
\bea
g(t,x)=\bar x\, (2T-t) +\frac{x-\bar x}{\kappa}(1-e^{-\kappa\, (2T-t)})
\eea
and it can be checked explicitly that  the general condition (\ref{dg}) is satisfied.  It follows 
\bea
d g(t,x)=-x\, dt +\frac{\sqrt{\eta}}{\kappa}(1-e^{-\kappa(2T-t)})\,dZ_t
\eea

Also the gain $g(s,x_s)$ is normally distributed, with conditional mean $M(s)$ and conditional variance $\Sigma^2(s)$  given by 
\bea\nn\label{gstat}
M(s)&:=&\mathbb{E}\,\left[ g(s,x_s)|x_t=x\right]=\bar x\, (2T-s) +\frac{x-\bar x}{\kappa}(1-e^{-\kappa\, (2T-s)}) e^{-\kappa (s-t)}\\
\Sigma^2(s)&:=&\text{Var}\,\left[ g(s,x_s)|x_t=x\right]=\frac{\eta}{2\kappa^3}(1-e^{-2\kappa(s-t)})(1-e^{-\kappa\,  (2T-s)})^2
\eea

\section{HJB equation with general price impact function}\label{genp}
In this appendix we write down the HJB equation for the case with a  general temporary impact function 
$ K \int |u_s|^p\, ds$ with $p>1$. The objective function (\ref{vex}) is generalized to the following function
\bea\label{vexp}
V(t,x,q) &=& \min_{\{u_s | s \in (t, T) \}} \mathbb{E}\left[\left. \int_t^T \left(C|u_s|- g(s,x_s) u_s\right) ds \right.\right. \nonumber \\ 
&&\left.\left. + K \int_t^T |u_s|^p ds +\frac{1}{2} \lambda \nu \int_t^{2T} (q_s-\bar q)^2 ds \right| q_t = q, x_t = x \right] \nonumber
\eea
which satisfies the HJB equation
\bea\label{hjbb}
\hat D_{t,x}\cdot V+ \frac{1}{2} \lambda \nu (q-\bar q)^2 + \min_u \left[ C|u| + K |u|^p + \left(\frac{\partial V}{\partial q} - g \right) u\right] = 0
\eea

As in the quadratic case discussed in the main text, due to the minimization  respect to $u$, the solution is divided into three trading regions: 
\begin{enumerate}
\item $g>C+\frac{\partial V}{\partial q}$. In this case $u>0$, so we \emph{ buy}:
\bea
u = \left(\frac{1}{p K }\right)^{\frac{1}{p-1}}  \left(g-C- \frac{\partial V}{\partial q}\right)^{\frac{1}{p-1}}
\eea
\item  $g<-C+\frac{\partial V}{\partial q}$. In this case $u<0$, so  we \emph{ sell}: 
\bea
u =-\left(\frac{1}{p K }\right)^{\frac{1}{p-1}}  \left(-g-C +\frac{\partial V}{\partial q}\right)^{\frac{1}{p-1}}
\eea
\item  $-C+\frac{\partial V}{\partial q}\le g \le C+\frac{\partial V}{\partial q}$.  In this case we \emph{ don't trade} ($u = 0$).
\end{enumerate}
The trade rate  $u$ is concisely written as:
\bea
u= \left(\frac{1}{p K }\right)^{\frac{1}{p-1}}  \left(g-C- \frac{\partial V}{\partial q}\right)_+^{\frac{1}{p-1}}-\left(\frac{1}{p K }\right)^{\frac{1}{p-1}}  \left(-g-C +\frac{\partial V}{\partial q}\right)_+^{\frac{1}{p-1}} 
\eea
and the HJB equation (\ref{hjbb}) can be written as:
\bea
\hat D_{t,x} \cdot V + \frac{1}{2} \lambda \nu (q - \bar q)^2 - K (p-1) |u|^p = 0
\eea

\section{Zero volatility signal}
\label{detsig}

We consider the approximation where the HF signal $x_t$ has zero volatility, that is
\bea
dx(t)= \mu(t,x(t)) dt 
\eea
In this case, the optimization problem is completely deterministic and the optimal strategy, defined as the strategy that minimizes the objective function (\ref{vex}),   can be computed using the variational principle.  When the initial position is located in the buy or the sell zone,  it is optimal to trade toward the no-trading (NT) zone. We assume that we reach the NT zone before the market closes, and once we have entered the NT zone, we stay there until the end of the day. As we will show, this is a correct assumption for the  case of an exponential signal (or zero signal).  In general, however, the boundaries of the NT region depend on $x$ and hence one can go in and out of the region many times, also for a deterministic signal.

Under our assumptions,  if the initial position $q(t)=q$ is located in the sell region, we sell until  we reach $b_-$,   the threshold between the sell and the NT region. That is,  we sell until a time $\hat t$ such that $q(\hat t )=b_-(\hat t, x(\hat t))$. Analogously, in the case where the initial position  is located in the buy zone, we buy until we reach  $b_+$, the threshold between the buy zone and the NT region.  That is,  until a stopping time $\hat t$ such that $q(\hat t)=b_+(\hat t, x(\hat t))$. We denote as  $V_-$ ($V_+$) the objective  function  (\ref{vex}) for the case where we start from the sell (buy) region.  Explicitly,
\bea\nonumber
V_\pm(t,x,q) &=& \min_{\{q(s) | s \in (t, T) \}} \left[ \int_t^{\hat t} \left(\pm C\dot q(s)- g(s) \dot q(s)+K\dot q(s)^2+\frac{1}{2} \lambda \nu (q(s, x(s))-\bar q)^2\right) ds  \right.\\&& \left.  +\frac{1}{2} \lambda \nu (2T-\hat t) ((q(\hat t) -\bar q)^2\right]\nonumber \\
&=& \min_{\{q(s) | s \in (t, T) \}} {\cal V}_\pm[q(s)]
\eea
where in the last line we defined ${\cal V}_\pm[q(s)]$, a functional of the trading trajectory $q(s)$, which satisfies the boundary condition $q(t)=q$ and $\dot q(s) := dq(s)/ds$. Our signal $x(s)$ is  deterministic with  boundary condition  $x(t)=x$, and the gain  is defined as $g(s, x(s))=\int^{2T}_s x(r) dr$, which is the deterministic limit of the gain defined in (\ref{gain}).
The optimal strategy is given by the trajectory $q(s)$ that minimizes ${\cal V}_\pm[q(s)]$, and can be derived  using the Euler-Lagrange variational principle. That is, we require ${\cal V}_\pm[q(s)]$ to be stationary with respect any admissible fluctuation around the minimizing trajectory.   The initial position is fixed, therefore we do not allow fluctuations at initial time, that is $\delta q(t)=0$.  However, in general  $\delta q (\hat t )\neq 0$, which will produce also equations of motion for boundary terms evaluated at time $\hat t$. We have
\bea\nn
\delta_{q(s)} {\cal V}_\pm[q(s)] 
&=&  \left( \lambda\nu (2T-\hat t ) \left(q({\hat t}) - \bar q \right) \pm C-g({\hat t, x(\hat t)})\right)\delta q({\hat t})\\  &&+  \int_t^{\bar t}\, ds\,\left( -2K\ddot{q}(s)+\dot g(s,x(s))+\lambda\nu (q(s)-\bar q) \right)\delta q(s)
\eea 
where we used $\dot q (\bar t)=0$. Indeed, formula (\ref{um}) shows that the trading rate goes to zero when we approach the NT zone.  It follows that the trajectory that minimizes the cost function is described by  two equations:
\bea\label{euler}\nonumber
0&=& \lambda\nu (2T-\hat t ) \left(q({\hat t}) - \bar q \right) \pm C-g({\hat t}, x({\hat t}))\\
 0&=& -2K\ddot{q}(s)+\dot g(s, x(s))+\lambda\nu (q(s)-\bar q)
\eea
We remind that $\hat t$ is the time when the position reaches the boundary of the trading region, therefore $q(\hat t)$ is equivalent to $b_-(\hat t, x(\hat t))$  ($b_+(\hat t, x(\hat t))$ ) if we started from the sell (buy) region.   It follows that the first equation can be written in the following way
\bea\label{detb}
b_\pm(\hat t, x(\hat t))=\bar q+ \frac{1}{\lambda \nu (2T-\hat t )}(g(\hat t, x(\hat t))\mp C)
\eea
which is an explicit expression for the boundaries of the trading regions.\footnote{In this formula $\hat t$ is the stopping time. However, changing the value of $K$,  $\hat t$ can span the whole trading interval $[t_\text{open},T]$, therefore we can replace $t\rightarrow \hat t$  in (\ref{detb}), and interpret the  (\ref{detb}) as the expression for the boundaries of the trading regions in the $(t,q,x)$ space.}   In order to integrate the second equation and solve the optimization problem,  we consider an explicit dynamics for the signal $x(t)$.

\subsection{Deterministic mean reverting HF signal}
We start considering  a deterministic mean reverting signal of the form 
\bea
dx(t)= -\kappa x(t) dt
\eea 
which can be integrated to 
\bea
x(s)=x e^{-\kappa (s-t)}
\eea
for $s>t$, and the associated gain is given by
\bea
g(s, x(s))=\frac{x}{\kappa}(e^{-\kappa(s-t)}-e^{-\kappa(2T-t)})
\eea
The second equation can be integrated  imposing the initial condition $q(t)=q$. It results
\bea\label{qs}
q(s)=x\,\frac{ e^{-A (s-t)}-e^{-\kappa(s-t)}}{2K \kappa^2-\lambda \nu}+ a\, e^{s A}(1-e^{-2(s-t)A}) +\bar q\, (1-e^{- A(s-t)})+ q\, e^{- A(s-t)} 
\eea
where  $A=\sqrt{\frac{\lambda \nu}{2 K}}$ and $a$ is an integration constant to be determined.  The stopping time $\hat t$ and the integration constant $a$ are fixed by     the following system:
\bea\nonumber\label{eom}
\dot q(\hat t)&=&0\\
q(\hat t)&=&b_\pm(\hat t, x(\hat t))
\eea
where in the second equation we take $b_-$ if we start from the sell zone and  $b_+$ if we start from the buy zone. 
This system can be solved numerically. In figure \ref{fig:det_mean} we plot the solution (\ref{qs}) when the initial position is in the sell zone, showing that the result is in perfect agreement  with a numerical optimization obtained with quadratic programing. We consider various values for the constant $K$, and as expected, as $K \rightarrow 0$ we go instantaneously to one of the boundaries of the NT zone (if we are out).  As $K\rightarrow \infty$ we trade slowly towards the boundary.
\myfig{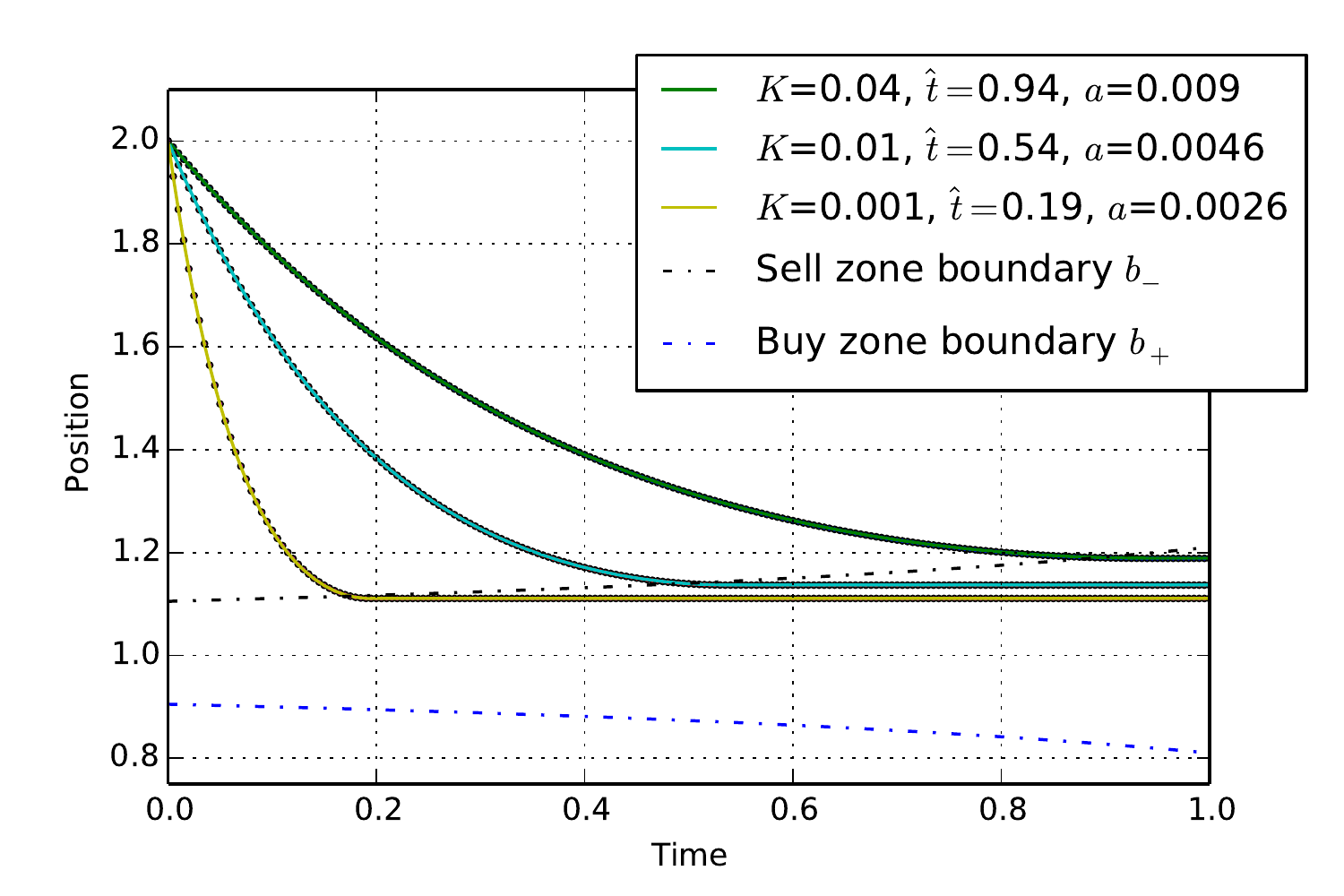}{0.8}{\small  In this figure we plot the optimal trading trajectory for the case where the initial position is in the sell zone. The continuos lines show the result (\ref{qs}) for several values of the constant $K$. The stopping time $\hat t$ and the integration constant $a$ are also reported in the legend. The dots show the results of a numerical optimization using quadratic programing.  The dashed lines show the boundaries $b_\pm$. The relevant parameters are chosen as follows:   $\nu=0.01$, $\lambda=50$, $C=0.1$, $\bar q=1.0$ and the mean reversion scale of the HF signal is $20$ mins. }{det_mean}

\subsection{General price impact function}
The variational principle can be applied also in the presence of a general price impact function. In particular, the equations (\ref{euler}) generalize to
\bea\nonumber
0&=& \lambda\nu (2T-\hat t ) \left(q({\hat t}) - \bar q \right) \pm C-g({\hat t}, x(\hat t ))\\
 0&=& \mp pK  \frac{d}{ds}(\pm\dot{q}(s))^{p-1}+\dot g(s,x(s))+\lambda\nu (q(s)-\bar q)
\eea
We notice that the first equation is equivalent to the first equation in (\ref{euler}). This implies that, for a deterministic signal, the boundaries of the trading regions are the same for any impact function.

\section{Market orders with quadratic costs}\label{quadcost}
In the case when we only consider quadratic costs (that is, $C = 0$) and market orders, the optimal trading rate is no longer discontinuous:
\be
u = \frac{1}{2K} \left(g-\frac{\partial V}{\partial q}\right)
\ee
The HJB equation then simplifies to:
\be
\label{qhjb}
D_{t,x} \cdot V + \frac{1}{2} \lambda \nu (q - \bar q)^2 - \frac{1}{4K} \left(g-\frac{\partial V}{\partial q}\right)^2 = 0
\ee
with the usual boundary conditions
\be
V(T,x,q)  = \frac{1}{2} \lambda \nu T (q - \bar q)^2
\ee
One can solve (\ref{qhjb}) by making the following quadratic ansatz:\footnote{Solutions of HJB equations with a polynomial structure are described in e.g. \cite{westray}.}
\be
V(t,x,q) = V_0(t,x) + V_1(t,x) (q-\bar q) + V_2(t,x) (q-\bar q)^2
\ee
which leads to the following PDEs
\bea
\hat D_{t,x} \cdot V_2 + \frac{1}{2} \lambda \nu - \frac{1}{K} V_2^2 &=& 0\\
\hat D_{t,x} \cdot V_1 + \frac{1}{K} V_2 (g - V_1) &=& 0\\
\hat D_{t,x} \cdot V_0 - \frac{1}{4K} (g - V_1)^2 &=& 0
\eea
with the boundary conditions:
\bea
V_2(T,x) &=& \frac{1}{2} \lambda \nu T \\
V_1(T,x) &=& 0 \\
V_0(T,x) &=& 0 
\eea

Using Feynman-Kac Theorem, it is easy to write down the general solution:
\bea
V_2(t) &=&  K A \frac{\tanh((T-t) A)+TA}{1+TA\tanh((T-t)A)}\\
V_1(t,x) &=& \frac{1}{K} \mathbb{E}\left[\left. \int_t^T e^{-\frac{1}{K} \int_t^s V_2(s') ds'} V_2(s) g(s,x_s) ds\right| x_t = x\right] \\
V_0(t,x) &=& -\frac{1}{4 K} \mathbb{E}\left[\left. \int_t^T \left(g(s,x_s) - V_1(s,x_s)\right)^2 ds\right| x_t = x\right]
\eea
where $A=\sqrt{\frac{\lambda \nu}{2 K}}$.

\section{$K \rightarrow \infty$ limit}
\label{kinf}

In the large $K$ limit,  we consider the following ansatz  for the objective function $V$:   
\bea\label{vkexp}
V=V_0+\frac{1}{K} V_1+\frac{1}{K^2} V_2+\ldots
\eea
that is we write $V$ as an expansion in $1/K$.    Plugging the ansatz (\ref{vkexp}) in the HJB equation (\ref{hjbm})  and collecting terms of the same order in  $1/K$, we can write down an equation for any term in the expansion (\ref{vkexp}), and solve the equation  (\ref{hjbm}) iteratively. We describe this prescription in details for the first two terms $V_0$ and $V_1$.

The leading order term $V_0$ satisfies the following equation

\bea\label{vz}
\hat D_{t,x} \cdot V_0 + \frac{1}{2} \lambda \nu (q - \bar q)^2= 0
\eea
with boundary condition 
\bea\label{bvz}
V_0(T,x,q) = \frac{1}{2} \lambda \nu T (q-\bar q)^2
\eea
The solution to equation (\ref{vz}) with boundary condition  (\ref{bvz})  is given by 
\bea
V_0= \frac{1}{2} \lambda \nu (2T-t) (q- \bar q)^2
\eea
which is the approximate solution given in (\ref{apx}).

The next to leading order term $V_1$ satisfies the following equation 
\bea\label{vo}
\hat D_{t,x} \cdot V_1-\frac{1}{4}  \left(g-C-\frac{\partial V_0}{\partial q}\right)^2_+-\frac{1}{4}  \left(-g-C+\frac{\partial V_0}{\partial q}\right)^2_+  = 0
\eea
with boundary condition 
\bea\label{bvo}
V_1(T,q,x) =0
\eea
The solution to equation (\ref{vo}) with boundary (\ref{bvo}) can be obtained using the Feynman-Kac formula
\bea\label{ofk}\nonumber
V_1(t,q,x) &=&-\frac{1}{4 } \mathbb{E}\,\left[\left. \int_t^T\, ds\,  \left( \left(g(s,x_s)-C-\frac{\partial V_0(s)}{\partial q}\right)^2_++ \left(-g(s,x_s)-C+\frac{\partial V_0(s)}{\partial q}\right)^2_+  \right) \right| x_t = x \right]\\
\eea
where the process $x_t$ is the general stochastic process in (\ref{gp}).   

\subsection{Boundaries}

The boundaries $b_\pm$ are defined by the equations (\ref{beq}). In the current perturbative scheme, they  are expanded as
\bea\label{bk}
b_\pm=b_{\pm}^{(0)} + \frac{1}{K}b_{\pm}^{(1)} + \frac{1}{K^2}b_{\pm}^{(2)}+\ldots
\eea
Using the expansions for $V$ (\ref{vkexp}) and the expansion for $b_\pm$ (\ref{bk}), the equation (\ref{beq}) is written as
\bea
g\mp C=\left.\frac{\partial V_0}{\partial q}\right|_{q=b^{(0)}_{\pm}}+\frac{1}{K}\left(\left.\frac{\partial V_1}{\partial q}\right|_{q=b^{(0)}_{\pm}}+\left.\frac{\partial^2 V_0}{\partial q^2}\right|_{q=b^{(0)}_{\pm}}\,\, b_{\pm}^{(1)} \right) +\ldots
\eea
which can be solved order by order in the $\frac{1}{K}$ expansion.  The leading order term $b_{\pm}^{(0)} $ is given in equation (\ref{bexp}).  The next to leading order term $b_{\pm}^{(1)}$ is given by 
\bea\label{bo}
b_{\pm}^{(1)}&=&-\frac{\left.\frac{\partial V_1}{\partial q}\right|_{q=b^{(0)}_{\pm}}}{\left.\frac{\partial^2 V_0}{\partial q^2}\right|_{q=b^{(0)}_{\pm}}}
\eea
In the case where  $g(s,x_s)$ is normally distributed    with conditional mean $M(s):=\mathbb{E}\,\left[ g(s,x_s)|x_t=x\right]$ and conditional variance $\Sigma^2(s):=\text{Var}\,\left[ g(s,x_s)|x_t=x\right]$, it is possible to evaluate the expectations in (\ref{ofk}) and compute $\frac{\partial V_1(t,q,x)}{\partial q}$.  \footnote{When $x_t$  is an Ornstein-Uhlenbeck mean reverting process,  $g(s,x_s)$ is normally distributed and the  expressions for mean and variance are given in    (\ref{gstat}).} We obtain
\bea\nonumber
\frac{\partial V_1(t,q,x)}{\partial q}&=&\frac{1}{2  }\int_t^T\, ds\,  \frac{\partial^2 V_0(s)}{\partial q^2}\,\Sigma(s)\, \left(\frac{e^{\frac{-\Lambda_+^2(s)}{2}}}{\sqrt{2 \pi}}-\frac{e^{\frac{-\Lambda_-^2(s)}{2}}}{\sqrt{2 \pi}}-\Lambda_+(s)(1-\Phi(\Lambda_+(s)))-\Lambda_-(s)\Phi(\Lambda_-(s))\right)\\
\eea
where 
\bea
\Lambda_{\pm}(s)=\frac{\pm C+\frac{\partial V_0(s)}{\partial q} - M(s)}{\Sigma(s)}
\eea
and $\Phi$ is the cumulative function of  the unit normal distribution function.  The conventions are chosen so that the terms depending on $\Lambda_-$ encode the contribution of the sell region and the terms with $\Lambda_+$,  are the contribution of the buy zone.

\end{document}